\documentclass[twocolumn,aps,pre,showpacs,amsmath,amssymb,floatfix]{revtex4-1}
\usepackage{graphicx}
\usepackage{dcolumn}
\usepackage{bm}
\usepackage[usenames,dvipsnames,svgnames,table]{xcolor}
\usepackage{wasysym}
\usepackage{pifont}
\usepackage[normalem]{ulem}

\begin{document}

\title[Sample title]{Casimir effect between pinned particles in two-dimensional jammed systems}

\author{Juan-Jos\'{e} Li\'{e}tor-Santos}
\author{Justin C. Burton}%
\email{Author to whom correspondence should be addressed:\newline justin.c.burton@emory.edu}
\affiliation{ 
Department of Physics, Emory University, Atlanta, GA 30033, USA
}%

\date{\today}

\begin{abstract}
The Casimir effect arises when long-ranged fluctuations are geometrically confined between two surfaces, leading to a macroscopic force. Traditionally, these forces have been observed in quantum systems and near critical points in classical systems. Here we show the existence of Casimir-like forces between two pinned particles immersed in two-dimensional systems near the jamming transition. We observe two components to the total force: a short-ranged, depletion force and a long-ranged, repulsive Casimir-like force. The Casimir-like force dominates as the jamming transition is approached, and when the pinned particles are much larger than the ambient jammed particles. We show that this repulsive force arises due to a clustering of particles with strong contact forces around the perimeter of the pinned particles. As the separation between the pinned particles decreases, a region of high-pressure develops between them, leading to a net repulsive force.

\end{abstract}
                              

\maketitle

\section{Introduction}
One of the most astonishing and counterintuitive ideas in modern physics is that the vacuum is not empty. Quantization of the electromagnetic fields leads to a non-zero ground-state energy density \cite{rugh,zelda}, which can have physical manifestations. In 1948, Hendrik Casimir predicted that adjacent conducting surfaces can confine the spectrum of zero-point modes, leading to a mutual attraction \cite{casimir}. Since then there have been a number of theoretical and experimental developments confirming this prediction, as well as identifying new electromagnetic, Casimir phenomena in various systems \cite{Klimchitskaya2009,Bordag2001,romeo,lam1,lam2,Sushkov2011}, including the onset of repulsive long-ranged forces when the boundary condition for ideal conducting plates is relaxed \cite{lif1,lif2,capasso}. 

While the original Casimir force is of purely quantum origin, there are analogous interactions that arise from the confinement of fluctuations in classical systems \cite{kardar,Balibar2005,Gambassi2009}. Perhaps the best known example of these is the critical Casimir force described by Fisher and de Gennes \cite{fish}. A binary liquid mixture close to the critical point experiences large-scale concentration fluctuations in the homogeneous phase. The fluctuations occur over a maximum correlation length which diverges as the system approaches the critical point. When objects are immersed in this mixture, they confine the fluctuations within the fluid between their surfaces, giving rise to a pairwise interaction which depends on the boundary condition at the surface \cite{Hanke1998,Gambassi2011}. For a single particle, the first direct experimental measurements verified theoretical predictions \cite{bechinger}. This interaction can also lead to aggregation and self-assembly of colloidal particles \cite{Bonn2009,Gambassi2010,Vasilyev2014,Edison2015}, a rich problem which includes many-body interactions \cite{Hobrecht2015,Paladugu2016}.


Critical Casimir forces are a universal feature of second-order phase transitions \cite{Gambassi2009,Krech1994,Gambassi2011}. They were first observed in thin $^{4}$He films near the superfluid $\lambda$-point \cite{Garcia1999}, and have also been observed in thin liquid films suspended on an immiscible liquid near the critical mixing point \cite{Fukuto2005,rafai}. In all cases, the force depends on the surface-surface separation, the geometry of the interfaces, and more importantly, on the boundary condition for the order parameter. Symmetric boundary conditions almost always give rise to an attractive force, and asymmetric boundary conditions lead to repulsive forces. 

Casimir forces have strengths that are proportional to the driving energy of the fluctuations, and thus proportional to $\hbar$ in quantum systems and temperature in classical systems. However, Casimir-like forces have recently been identified in a number of nonequilibrium systems as a result of the confinement of fluctuations \cite{Hanke2013,Brito2007}. In most of these cases, the fluctuations arise from energy input into the system either by an external field or by the individual particles \cite{grano2}. Examples include driven acoustic noise in a gas \cite{Larraza1998}, soft modes in polymer melts \cite{Obukhov2005}, temperature gradients in liquids \cite{Kirkpatrick2015,Najafi2004}, active matter composed of swimming particles \cite{Ray2014}, and hydrodynamic fluctuations in driven granular fluids \cite{catt,duncan,zuri,villa,reza}. In many of these examples, the Casimir-like forces do not obey a simple scaling function in accordance with Fisher and de Gennes' original argument \cite{fish}. In non-equilibrium systems, the boundary conditions can depend on the order parameter, and the spectrum of fluctuations are not well-defined, in contrast to thermal and quantum systems.

As opposed to classical thermodynamic phase transitions where temperature plays a pivotal role, the jamming transition in granular materials is mainly governed by changes in the density \cite{liu}. In the simplest scenario, a jammed system is composed of frictionless spheres with finite-ranged interactions at $T=0$ (zero kinetic energy).  As the density changes, the system reaches a special point, called point $J$, where the spheres begin to touch one another and the system develops a finite pressure \cite{jamm1,corey}. The physics at point $J$ shares many similarities to second-order phase transitions \cite{Drocco2005,Head2009}, yet it is distinctly different. It is a rare example of a random first-order transition in which there are several length scales that diverge as the critical density of the system is approached \cite{Biroli2007,biroli,goodrich2,goodrich3,Reichhardt2014}. 

\begin{figure}
\includegraphics[width=.47 \textwidth]{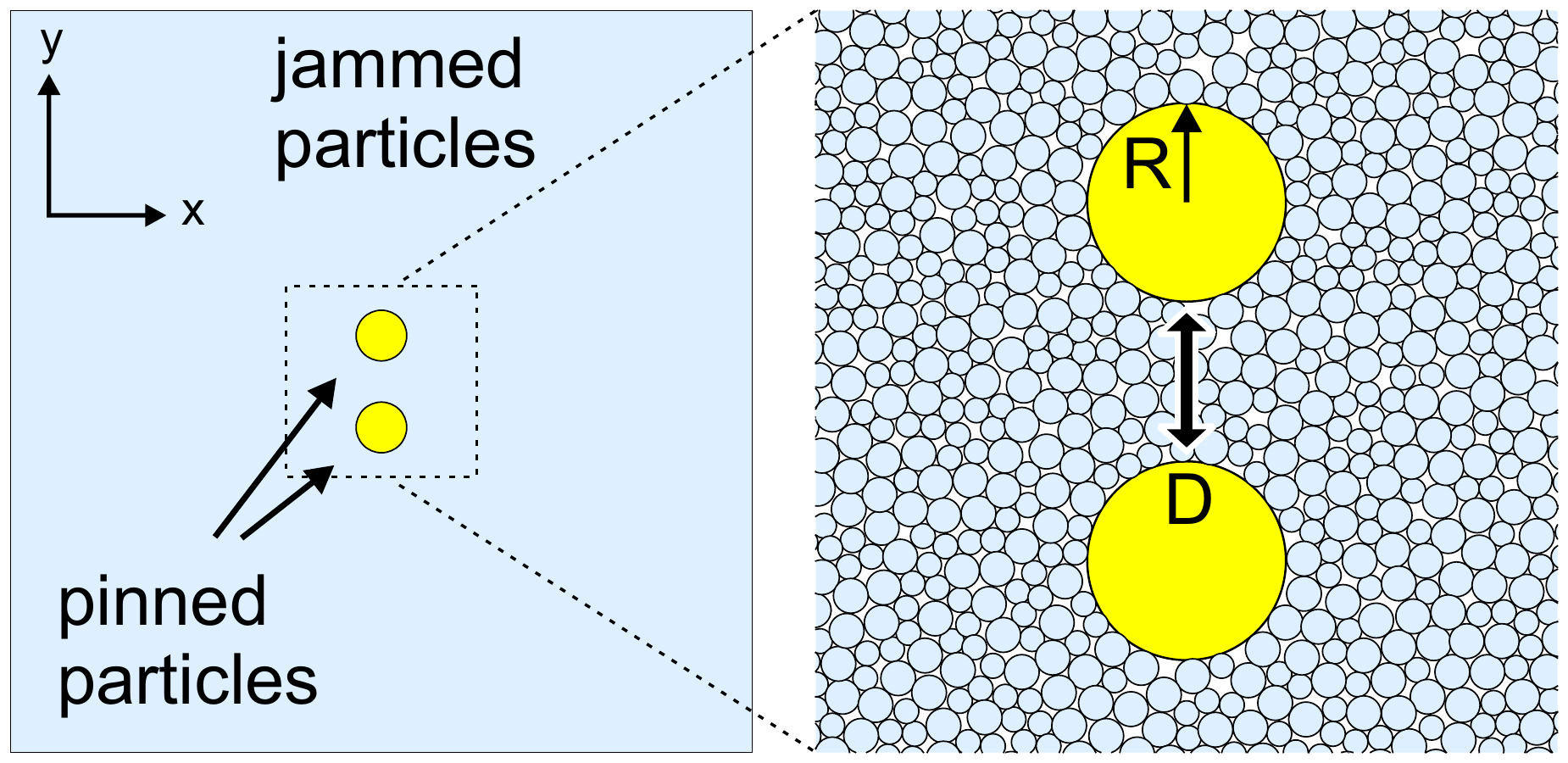}
\caption{\label{cartoon}Schematic showing a typical simulation. Two pinned particles (yellow) of radius $R$ are held in place while a sea of bidisperse disks are quenched around them. The distance between their surfaces is $D$. In equilibrium, the pinned particles retain a net force, which is equal and opposite.}
\end{figure}

Here we report simulations which show the existence of Casimir-like forces in systems composed of frictionless particles near the $T=0$ jamming transition in two dimensions. The particles interact through a finite-ranged repulsive force, so that there is a well-defined isostatic point as the density is varied. At this isostatic point, the number of mechanical contacts between the particles is equal to the number of motional degrees of freedom. Two equal-sized particles are pinned while the rest of the system, composed of $N$ particles of two sizes, are quenched to mechanical equilibrium, as shown in Fig.\ \ref{cartoon}. Because the pinned particles are fixed in place, they retain a net force, which has two distinct components. The particles experience a short-ranged depletion force which depends on the details of the pair-distribution function, and a long-ranged repulsive force. This long-ranged force increases in magnitude near the jamming transition, as does the fluctuations of this force between different particle configurations. 

We identify this long-ranged force as a Casimir-like force induced by fluctuations in systems near jamming. The specific dependence of the force on particle separation will be determined by the nature of the fluctuations which give rise to the force. In granular systems, it is well-known that forces are spatially heterogeneous and are clustered in into linear chains \cite{Majmudar2005,Giusti2016}. The distributions of the strength and length of these force chains have been studied \cite{Sanfratello2011,Liu1995,Peters2005,Corwin2005}. These studies, and many others, have shown that both strong and long force chains are exponentially rare, and depend on the state and history of the system. For example, shearing a granular system in one direction will make force chains preferentially align to resist the applied shear, and can induce jamming in unjammed systems \cite{Bi2011}.

While it is not clear how confinement, packing fraction, and system preparation affects the distribution of force chains, our results show that near the jamming transition, the Casimir-like force follows a universal form which scales as $B-A\Delta\phi^{1/2}$, where $\Delta\phi=\phi-\phi_c$ is the distance to the isostatic packing fraction $\phi_c$, and the parameters $B$ and $A$ decrease as the pinned particles move further apart.  In our simulations, the system is at finite pressure, so that $\phi$ is strictly greater than $\phi_c$. This Casimir-like force is distinctly different than other Casimir-like forces reported in granular systems \cite{catt,duncan,zuri,villa,reza}. In the latter, hydrodynamic fluctuations are induced by external mechanical excitation, so that there is significant kinetic energy in the system ($T>0$). In our simulations, the Casimir-like force is determined by the critical behavior and inherent fluctuations associated with the $T=0$ jamming transition \cite{Goodrich2014b,Sanfratello2011}.

The origin of this force lies in the distribution of particles with large forces in systems near jamming, and the coordination number of the jammed particles near the boundary of the pinned particles. We find that the inclusion of pinned particles in a jammed system reduces the mean coordination number of the entire system. This reduction is mostly localized near the boundary of the pinned particles. In addition, near jamming, the largest contact forces occur between particles with the minimum number of contacts, which is $z=3$ in two dimensions. These two properties result in the region between the pinned particles having less than the mean number of contacts, and large contact forces, which push the pinned particles apart. Although not explicitly investigated here, we also show how our results may be generalized to higher dimensions.

\section{Methods}
\label{methods}

Our simulations consist of $N$ particles that interact through a finite-ranged, repulsive potential \cite{corey}:

\begin{equation}
\label{pot}
V(r_{ij})= 
\begin{cases} 
K\dfrac{\epsilon}{\alpha}\left(1-\dfrac{r_{ij}}{\sigma_{ij}}\right)^{\alpha}
 & \text{for $r_{ij}<\sigma_{ij}$,}
\\
0 &\text{for $r_{ij}>\sigma_{ij}$,}
\end{cases}
\end{equation}

\noindent
where $\epsilon$ is the energy scale of the interaction, $r_{ij}$ is the distance between the centers of particles $i$ and $j$, $\sigma_{ij}=\sigma_i+\sigma_j$ is the sum of the particle radii, and $\alpha$ defines the type of interaction. Unless otherwise noted, all simulations use $\alpha=2$ (harmonic interactions), however, we obtain similar results with other values such as $\alpha=5/2$ (Hertzian interactions). In order to prevent crystallization, we use a binary mixture of particles with radii $\sigma$ and $1.4\sigma$, with equal numbers of both particles. For these $N$ particles, the constant $K=1$. In addition to these $N$ particles, we include two particles of radius $R$ separated by a distance $D+2R$, so that the surface-surface separation is $D$ (Fig.\ \ref{cartoon}). These particles are pinned in place, and not allowed to move. For large values of $R$, Eq.\ \ref{pot} 
produces a very soft potential so that the free particles easily penetrate deep into the boundary of the pinned particles. To avoid this, for particles interacting with the pinned particles, we set $K=(R+1.4\sigma)^{\alpha}/(2.8\sigma)^{\alpha}$. This way the force between pinned particles and free particles is finite in the limit $R\rightarrow\infty$. Near the jamming transition ($\Delta\phi=0$) where the Casimir-like force dominates, our results are insensitive to the choice of $K$ since the overlap distance is much less than both $R$ and $\sigma$.

Each simulation is initiated by fixing the positions of the pinned particles, and placing the remaining $N$ particles down at random in a square box with sides of length $L$ with periodic boundary conditions on all sides. The free particles are then instantaneously quenched to the nearest equilibrium configuration using the FIRE algorithm \cite{fire,Burton2016}. The algorithm is terminated when all the kinetic energy has been removed from the system. Our condition for equilibrium was chosen so that the magnitude of any net force on the free particles is less than $10^{-14}\times\epsilon/\sigma$. Once the system reached equilibrium, we measured both the $x$ and $y$ components of the residual force on the pinned particles ($F_x$ and $F_y$). Since these two particles are the only fixed objects in the system, they must experience equal and opposite forces according to Newton's third law. Thus we report only the force on the upper particle, so a positive force in the $y$-direction corresponds to a net repulsive force between the two pinned particles. The net force components on the pinned particles will fluctuate between different quenched systems, and can be positive or negative,  thus we report ensemble averages for the measured forces. Of special interest is the standard deviation of the force distribution in an ensemble, since this is indicative of the inherent force fluctuations in jammed systems, and will be discussed in section \ref{results}. 

For small values of $\Delta\phi$, the forces between particles are naturally smaller since there is less overlap. It is more relevant to study the net force normalized by the system pressure and the radius of the pinned particles, $F/(PR)$. The system pressure was measured at each packing fraction, as defined in reference \cite{corey}:
\begin{equation}
P=-\dfrac{1}{2L^2}\sum_{i<j}r_{ij}\dfrac{dV_{ij}}{dr_{ij}}.
\label{pres}
\end{equation}
Here the summation runs over all free and pinned particles in the system. All force measurements reported here will be normalized in this fashion. In addition, the effects of the initial placement of particles, as well as system annealing, will be discussed in the results.

\section{Results and Discussion} \label{results}

There are two distinct components of the force between pinned particles: a short-ranged force which depends on the local packing structure, and a long-ranged Casimir-like force. The relative strength of these two forces will depend on many length scales, such as $\sigma$, $R$, and $D$. In addition, there are multiple characteristic length scales in the jamming transition which diverge as $\Delta\phi\rightarrow 0$. These length scales are associated with the smallest rigid cluster that can be formed at a given $\Delta\phi$ \cite{goodrich2}, and the system's stability to transverse deformations \cite{goodrich3}. For the time being, we associate a general correlation length, $\xi$, with these diverging length scales and we will discuss its relation to both later in the results. Lastly, the size of the bounding box, $L$, will play an important role due to the prescribed periodic boundary conditions. Since the Casimir-like force is long-ranged, we can not ignore the contribution from image particles in neighboring simulation cells. 

In general, the normalized force between the two pinned particles will be a function of the ratio of these length scales:
\begin{align}
\left\langle\frac{F_y}{PR}\right\rangle=\Theta\left(\frac{D}{R},\frac{\sigma}{\xi},\frac{R}{\sigma},\frac{R}{L}\right).
\label{fulldep}
\end{align}
In practice, any linearly-independent combination of the ratios of the four length scales would do, however, we have chosen this combination because they are most relevant to our results. A particularly important limiting case, which will be discussed in section \ref{long_ranged}, is when $R/\sigma\rightarrow\infty$ and $R/L\rightarrow 0$. In this limit the details of the jammed medium between the pinned particles are less important, and effects due to the finite size of the simulation domain are negligible. In our simulations these finite-size effects can not be ignored, and will affect the data in different ways.

\begin{figure}[t]
\includegraphics[width=.47 \textwidth]{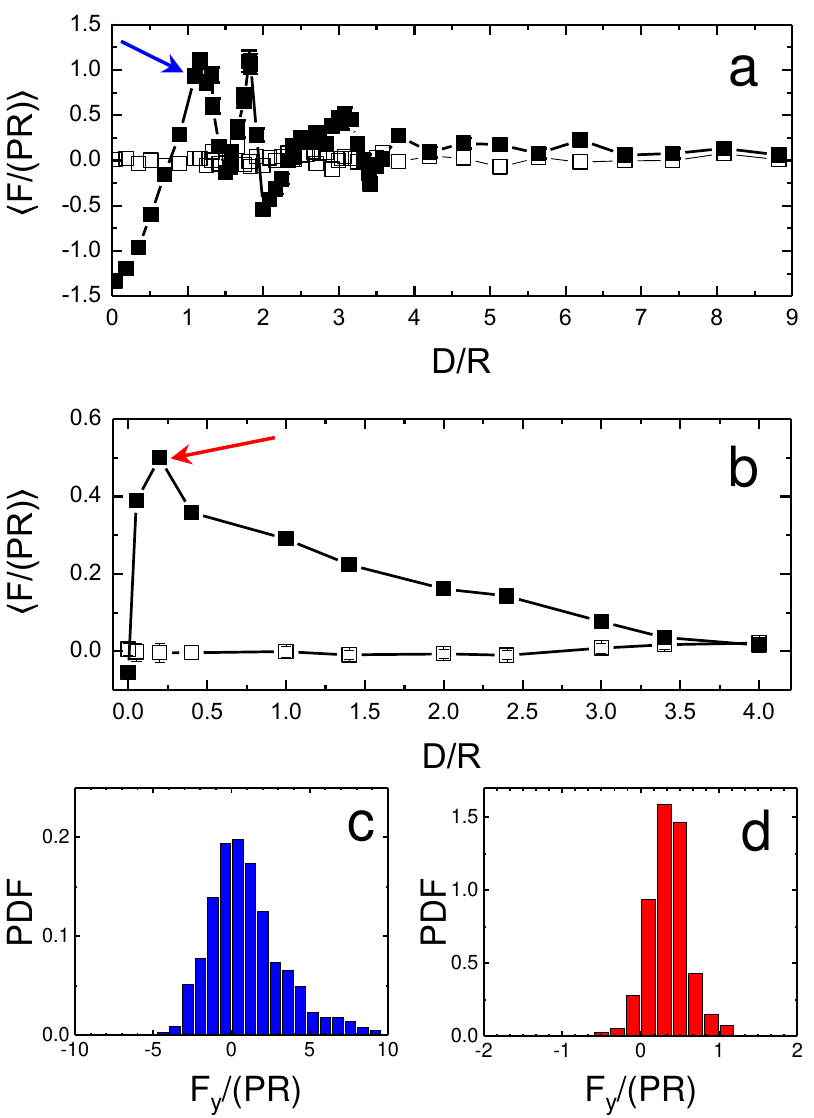}
\caption{\label{force_and_standev} (a-b) Components of the mean normalized force between the two pinned particles for two different values of $R/\sigma$: (a) $R/\sigma$ = 1.4, and (b) $R/\sigma$ = 21 ($L/\sigma\approx235$, $\phi = 0.855$, $F_x$ $\square$, $F_y$ $\blacksquare$). Each data point is the average of 200 systems with $N$ = 10,000 particles. PDFs of $F_{y}/(PR)$ are shown in panels (c) and (d), which correspond to the points indicated by the blue arrow in (a) and red arrow in (b).  }
\end{figure}

First, the basic dependencies of the residual force on $D/R$ and $R/\sigma$ are illustrated in Fig.\ \ref{force_and_standev}a-b, which shows the $x$ and $y$-component of the mean force ($\langle F_x\rangle$ and $\langle F_y\rangle$) for two different size ratios at $\phi$ = 0.855 ($\Delta\phi\approx$ 0.015). We observe that $\langle F_{x}\rangle=0$, as expected due to the symmetry of the geometry, as shown in Fig.\ \ref{cartoon}. However, $\langle F_y\rangle$ can vary significantly with particle size, $R/\sigma$, and particle separation, $D/R$. Small pinned particles can experience large positive and negative forces, which decay in amplitude with $D$, although $\langle F_y\rangle$ remains slightly positive for large $D$ (Fig.\ \ref{force_and_standev}a). For large pinned particles, the net force can be negative when the pinned particles are very close ($D/\sigma\sim 1$), as shown in Fig.\ \ref{force_and_standev}b. This is due to depletion. However, for larger values of $D$ there is only a repulsive force, which slowly decays as $D$ increases. 

We note here that by symmetry, $\langle F_y\rangle$ will be zero when the distance between the center of the pinned particles is equal to half the system size, which occurs when $D=D^*$, where $D^*$ is given by
\begin{align}
D^*\equiv\frac{L}{2}-2R.
\label{dstar}
\end{align}
This is because of the periodic boundary conditions: each pinned particle feels the same interaction from the image particles in neighboring, tessellated systems. This will be important when comparing systems with different values of $R/\sigma$, and will be discussed in detail. For smaller values of $D/R$, as $D\rightarrow 0$, eventually a point will be reached where only a few jammed particles (or none) can fit between the pinned particles. This depletion-like effect results in a net attractive force since little or no particles exist to push the pinned particles apart. This is true for both small and large values of $R/\sigma$, as shown in Fig.\ \ref{force_and_standev}a-b. 

Each data point in Fig.\ \ref{force_and_standev}a-b represents the average of 200 independent systems which have been quenched from a random configuration. Figure \ref{force_and_standev}c-d shows the distribution of forces obtained from these systems for two representative data points, as indicated by the blue and red arrows.  We found that the distributions, in general, are not symmetric. In addition, the mean is of the same order as the standard deviation. These characteristics are in contrast to Casimir-like forces in other non-equilibrium systems, such as driven granular gases, where the distributions are symmetric and the mean is less than 5\% of the standard deviation \cite{catt}. In the limits $R/\sigma\rightarrow\infty$ and $D/\sigma\rightarrow\infty$, the distributions will become very narrow due to the dramatic increase in possible jammed system configurations surrounding the pinned particles. 

\subsection{Short-Ranged Forces: $R/\sigma\sim 1$}

\begin{figure}[t]
\includegraphics[width=.47 \textwidth]{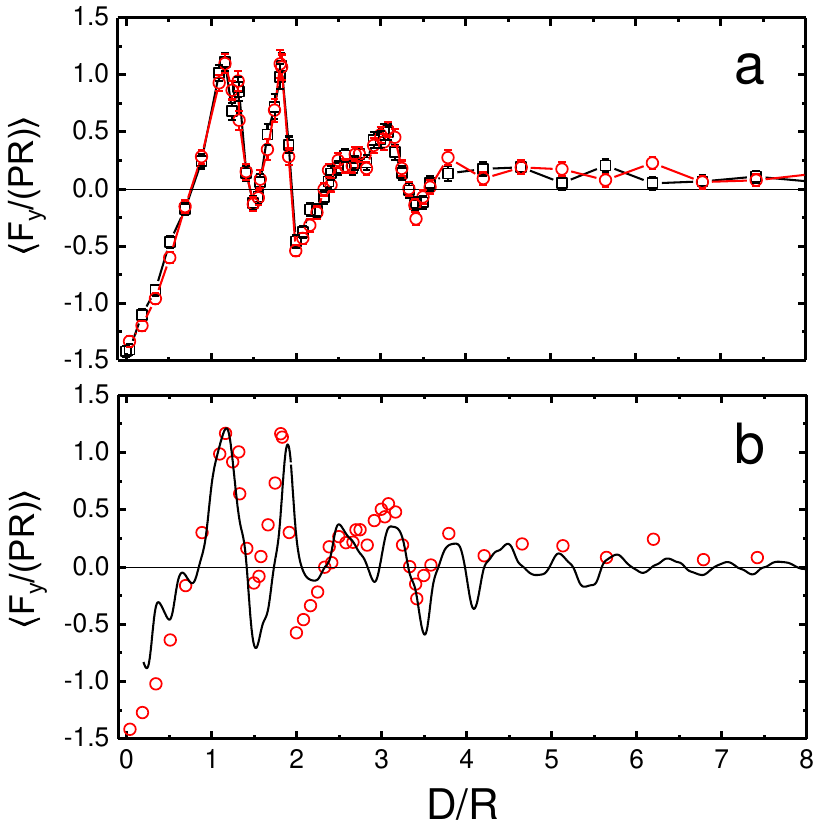}
\caption{\label{gofr_force} (a) Normalized force in the $y$-direction versus particle separation for two different system sizes ($R/\sigma$ = 1.4, $\phi = 0.855$. $N=1000$ {\color{black}$\square$}, $N=10,000$ {\color{red}$\fullmoon$}). Each data point represents the average of 800 systems. (b) Data with $N=10,000$ compared with the prediction from Eq.\ \ref{pmf}. }
\end{figure}

First we will focus on the behavior of $\langle F_y\rangle$ when $R/\sigma = 1.4$. Figure \ref{gofr_force}a illustrates this behavior for two system sizes, $N=1000$ and $N=10,000$. The curves are essentially identical, indicating that the important qualitative features of $\langle F_y\rangle$ are independent of the system size, so long as $D\ll L$. Near $D=0$, there is a net attraction between the pinned particles, as evidenced by an overall negative force that further fluctuates between negative and positive values as the distance increases. We attribute these initial fluctuations to a net depletion interaction between the pinned particles due to the amorphous-like structure of the jammed spheres. Similar phenomenology has been observed in experiments with colloidal particles \cite{arjun}. To further strengthen this point, we compare our force values with the result from the force obtained from the mean force potential, which we calculate using the pair correlation function \cite{mcquarrie}:
\begin{equation}
\langle F_y\rangle=k_BT\dfrac{g'(r)}{g(r)}.
\label{pmf}
\end{equation}
Here we have computed $g(r)$ by computing the probability of finding two particles with radii 1.4$\sigma$ at a given value of $D/R$. This method assumes an equilibrium, thermodynamic system with temperature $T$, which defines the fluctuating pathways between different system configurations. Our system configurations are obtained through a quench from $T=\infty$, thus we scaled the amplitude of the predicted force from Eq.\ \ref{pmf} in order to best fit our experimental curve (Fig.\ \ref{gofr_force}b). At this point, the quantitative meaning of the effective temperature used to scale the amplitude is unclear, however, the reasonable agreement between the functional form and the data indicates that the pair distribution function captures the essential features of $\langle F_y\rangle$ for small values of $D/R$. Oscillations in the predicted curve at small values of $D$ are likely due to the differentiation of discrete data from measured values of $g(r)$.

\begin{figure}[t]
\includegraphics[width=.47 \textwidth]{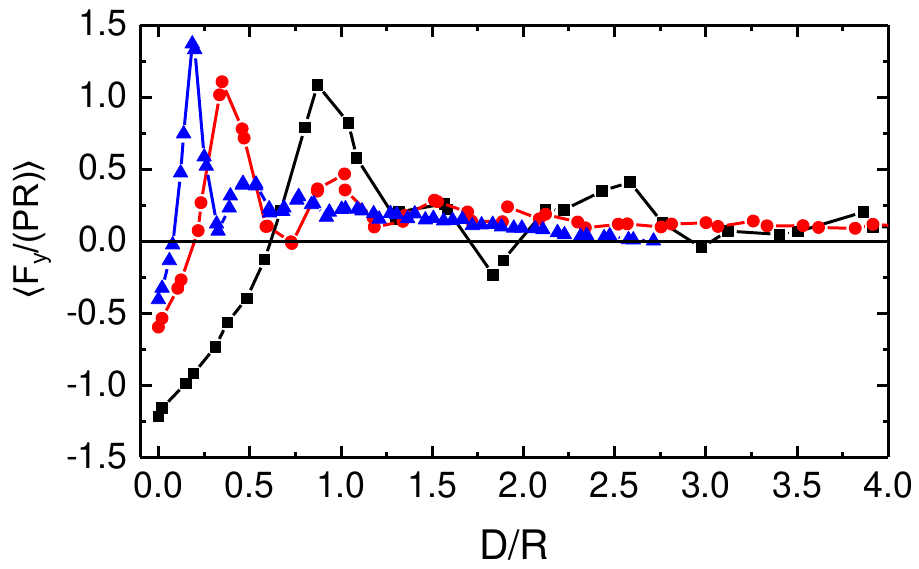}
\caption{\label{sizerat_force} Normalized force along the $y$-direction versus particle separation for different size ratios: ($\blacksquare$) $R/\sigma$ = 1.6 , ({\color{red}$\newmoon$}) $R/\sigma$ = 4, ({\color{blue}$\blacktriangle$}) $R/\sigma$ = 8. The packing fraction, $\phi = 0.89$, and the number of particles, $N=1000$, are constant. Each data point represents the average of 800 systems.}
\end{figure}

At larger distances, however, we find that the average force does not tend towards zero as we would expect from liquid-like ordering, rather it remains positive and therefore repulsive. This is indicative of a long-ranged Casimir component to the force, such as those that arise in binary mixtures \cite{fish,bechinger} and in granular systems \cite{catt,grano2}. In order to isolate this component of the force, we gradually increased the size of the pinned particles with respect to the jammed particles. For larger values of $R/\sigma$, the pinned particles have many neighboring particles and the net force does not depend as sensitively on the details of the local packing structure. However, the Casimir-like force becomes more prominent. This can be seen in Fig.\ \ref{sizerat_force}, which shows $\langle F_y \rangle$ for three different values of $R/\sigma$, keeping constant the packing fraction $\phi=0.89$ and the number of particles $N=1000$. As $R/\sigma$ increases, only the positive repulsive force remains, except at very small surface separations when $D/\sigma\sim1$, at which point the packing structure of jammed particles in the region between the pinned particles can strongly influence the mean force. For larger values of $D/R$, the force must tend to zero by the symmetry of the periodic boundary conditions, which can be seen here for $R/\sigma=8$.

\subsection{Long-Ranged Forces: $R/\sigma\gg 1$}
\label{long_ranged}

In order to better understand the Casimir-like component to the total force, we now turn out attention to systems where the pinned particles are significantly larger than the jammed particles ($R/\sigma\geq 20$). For these systems, we performed a detailed study of the dependence of the Casimir-like force on $\Delta\phi$, which controls the distance to the jamming transition. As we approach the critical point, we expect both the size and length scale of the fluctuations to increase \cite{Ellenbroek2006}. In order to obtain systems arbitrarily close to $\Delta\phi=0$, we began with a system at a high packing fraction, and slowly decompressed the system and monitored the decreasing pressure (Eq.\ \ref{pres}). 

\begin{figure}[t]
\includegraphics[width=.47 \textwidth]{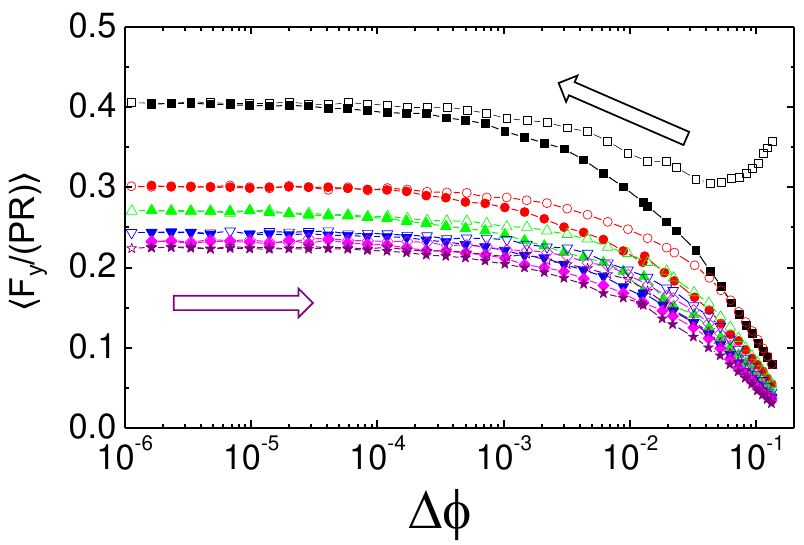}
\caption{\label{back_and_forth} Mean force in the $y$-direction between two pinned particle upon repeated decompressions and compressions for $D/R=1$ and $R/\sigma=21$. Each data point represents the average of 200 systems with $N$ = 10,000 particles each. Starting from $\Delta\phi=0.15$ ({\color{black}$\square$}), the systems are decompressed towards the jamming transition, then re-compressed. The final state of the system is lowest curve ({\color{Purple}$\bigstar$}). }
\end{figure}

The pressure scales with $\Delta\phi^{\alpha-1}$ as the jamming transition is approached \cite{corey}.  Systems were initially created at a packing fraction of $\phi$ = 0.99, and the packing fraction was decreased slowly in order to obtain a relationship between $P$ and $\phi$. During this process, $R$ was fixed and $\sigma$ was adjusted in order to change the overall packing fraction. Using Newton's root-finding method, we estimated the critical packing fraction where the pressure reaches zero. At each iteration, we reduced $\Delta\phi$ by 1/4, then re-quenched the system, thereby creating a series of packings ranging from $10^{-6}<\Delta\phi<0.15$. This algorithm had a significant effect on the Casimir-like force. Systems which are quenched to the nearest potential energy minimum from $T=\infty$ can rest in a rather high energy state. That is, any amount of annealing can relax the system considerably. This can be seen in Fig.\ \ref{back_and_forth}, which shows a series of decompressions and compressions for $D/R=1$ and $R/\sigma=21$. 

We started our systems at $\phi=0.99$ ($\Delta\phi\approx 0.15$), and then reduced the packing fraction in small steps while monitoring the net force on the pinned particles. Initially the force drops, and then begins to increase before reaching a plateau as $\Delta\phi\rightarrow 0$. We attribute the initial drop in the force to an annealing of the system during the initial decompression. The force decreases because free particles in between the two pinned particles are allowed to rearrange and relax. Upon compressing the system  back again to large values of $\Delta\phi$, the force drops monotonically. Further decompressions and compressions result in the same qualitative behavior, with the force approaching a steady value of $\langle F_y/(PR)\rangle\approx 0.23$ as $\Delta\phi\rightarrow 0$. Since multiple compressions and decompressions require significant computing time, data reported here has been decompressed, then compressed once (e.g. solid black squares in Fig.\ \ref{back_and_forth}). Further annealing of the systems would result in the same quantitative conclusions, just with smaller overall forces. 

\begin{figure}[t]
\includegraphics[width=.47 \textwidth]{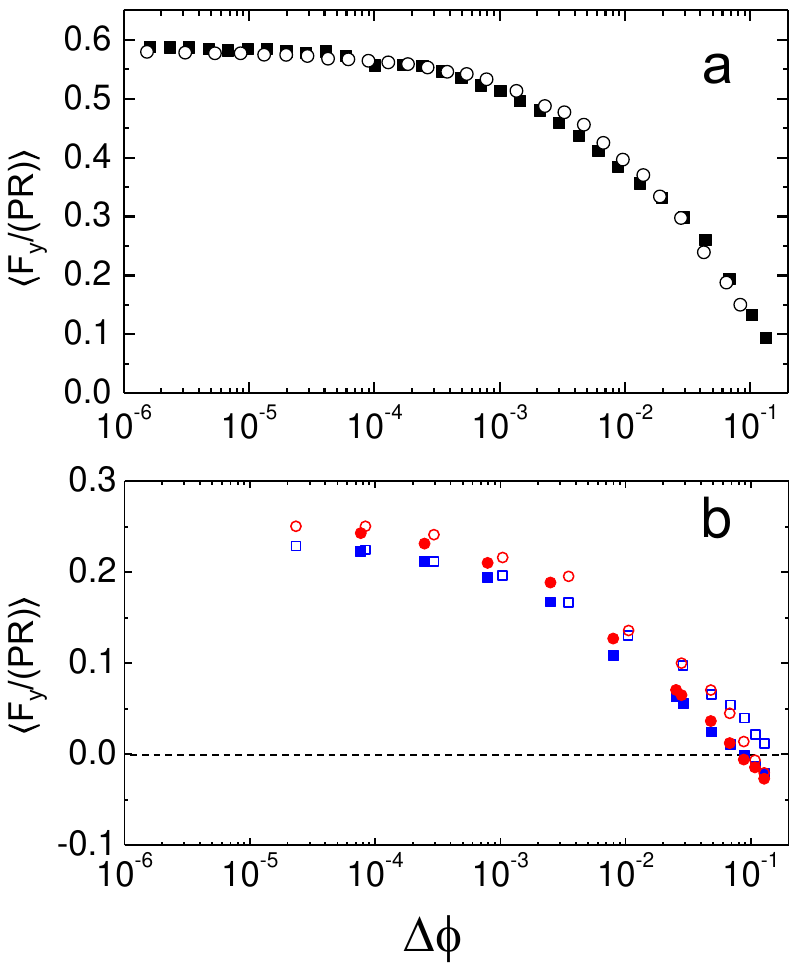}
\caption{\label{other_comp} (a) Comparison of the normalized force versus $\Delta\phi$ for both Harmonic ($\blacksquare$) and Hertzian ($\fullmoon$) interactions with $D/R=0.4$ and $R/\sigma=21$. (b) Data for two sequential sets of decompression and compression where all free particles are initially placed outside the boundary of the pinned particles.The order of the events is given by {\color{red}$\newmoon$}, {\color{red}$\fullmoon$}, {\color{blue}$\blacksquare$}, {\color{blue}$\square$}. All data points represent the average of 200 systems with $N$ = 10,000 particles each.}
\end{figure}

Figure \ref{other_comp}a shows that our results are independent of the type of interaction (the value of $\alpha$ in Eq.\ \ref{pot}). The mean force is shown for both harmonic and Hertzian interactions, with identical results. We also investigated the sensitivity to the initial placement of free particles prior to the quench. For most data, the particles were placed randomly, which means they could easily end up \emph{inside} the pinned particles, and then get pushed out during the quench. Alternatively, we  also placed the free particles in a random fashion only on the \emph{outside} of the pinned particles. This had the interesting effect of making the mean force negative (attractive) for large values of $\Delta\phi$. This is likely due to the fact that upon quenching, particles must be pushed into the region between the pinned particles, resulting in a lower average density and depletion-like effect. Nevertheless, upon approach to the jamming transition, the force becomes repulsive. This is shown in Fig.\ \ref{other_comp}b.

For critical Casimir forces, such as those that arise near classical critical points, there are two limiting regimes which have simple scalings in two dimensions for symmetric (Dirichlet) boundary conditions \cite{Hanke1998,Zubaszewska2013}. First, in the limit $R\ll D\ll \xi$, the Casimir force between two equally-sized particles scales as $D^{-5/4}$. In the opposite limit, $D\ll R\ll \xi$, where the Derjaguin approximation is valid, the force scales a $D^{-3/2}$. Thus the Casimir force may be expected to diverge as $D\rightarrow 0$. In practice, this divergence is limited by the assumption that the critical medium is a continuum, i.e. these scalings are valid only when $D$ and $R$ are much larger than the molecular length scale, an assumption which is violated as $D\rightarrow 0$. In our jammed systems, behavior near $D/R\rightarrow 0$ will be significantly affected by the size ratio $R/\sigma$, since this determines the degree to which the jammed particles can be considered a continuum. 

\begin{figure}[ht!]
\includegraphics[width=.47 \textwidth]{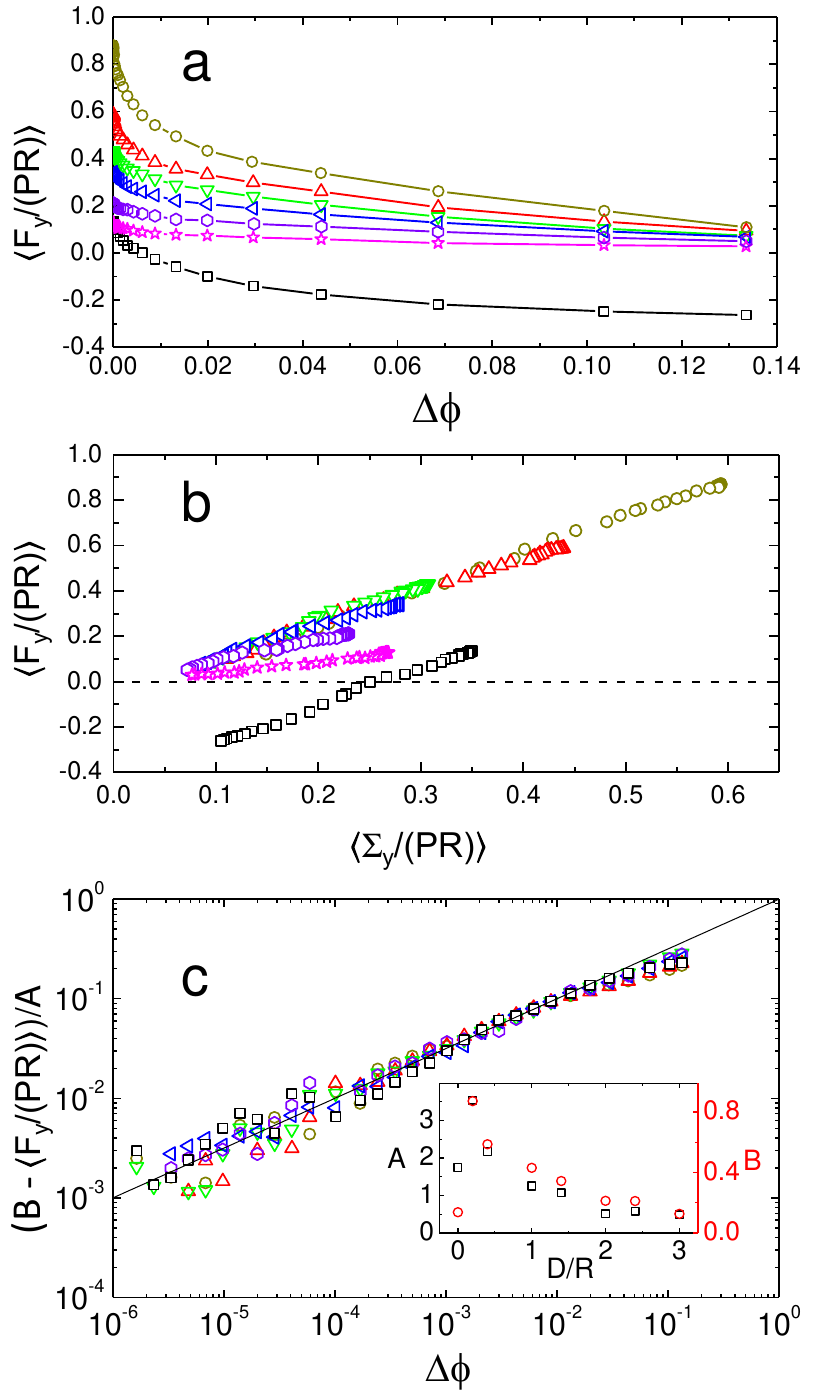}
\caption{\label{force_vs_phi} (a) Normalized force along the $y$-direction versus $\Delta\phi$ for different surface-to-surface distances: ($\square$) $\frac{D}{R}$ =0 ,  ({\color{Dandelion}$\fullmoon$}) $\frac{D}{R}$ = 0.2, ({\color{red}$\triangle$}) $\frac{D}{R}$ = 0.4, ({\color{green}$\triangledown$}) $\frac{D}{R}$ = 1, ({\color{blue}$\triangleleft$}) $\frac{D}{R}$ = 1.4, ({\color{violet}$\varhexagon$}) $\frac{D}{R}$ = 2.4, ({\color{CarnationPink} \ding{80}}) $\frac{D}{R}$ = 3. (b) Force versus the standard deviation of the force distribution for the data shown in (a). (c) The same data collapsed onto a universal functional form: $\langle F_y/(PR)\rangle=B-A\Delta\phi^{1/2}$. Each data point represents the average of 200 independent systems with $N$ = 10,000. The inset shows the fitting parameters $A$ and $B$ for each value of $D/R$. For all data, $R/\sigma$ = 21.}
\end{figure}

For jammed systems, the equivalent limit of $\xi\rightarrow\infty$ corresponds to $\Delta\phi\rightarrow 0$. Figure \ref{force_vs_phi} shows the approach to this limit at various values of particle separation $D/R$, with $R/\sigma$ = 21. In contrast to Fig.\ \ref{back_and_forth} and \ref{other_comp}, data in Fig.\ \ref{force_vs_phi}a is plotted on a linear-linear scale in order to illustrate the sharp increase in the mean force near $\Delta\phi=0$. When the pinned particles are touching each other at $D/R=0$, the mean force is negative for large $\Delta\phi$. This is due to the depletion effect, when $D/\sigma\lesssim 1$. However, as $\Delta\phi$ is decreased, the force sharply increases and becomes repulsive. This sharp increase is seen in all the data, but is weaker as the separation between the pinned particles increases. 

Figure \ref{force_vs_phi}b illustrates the relationship between the normalized mean force and the normalized standard deviation of the force distributions, such as those shown in Fig.\ \ref{force_and_standev}c-d. The relationship is always linear, although the slope and intercept vary with parameters such as $D/R$. The data indicates that the size of the fluctuations from system-to-system controls the maximum possible repulsive force. This is consistent with our expectation that Casimir-like forces arise from fluctuations in the system, where the fluctuations here are measured by the standard deviation in the net force over multiple jammed systems. 

Near $\Delta\phi=0$, all of the data is consistent with a universal functional form: 
\begin{align}
\left\langle \frac{ F_y}{PR}\right\rangle=B-A\Delta\phi^{1/2}.
\label{nearjam1}
\end{align} 
We fit all data points below $\Delta\phi=2\times10^{-3}$ to this form to determine the values of $A$ and $B$. A collapse of the data can be seen in Fig.\ \ref{force_vs_phi}c. The data follows a power law with an exponent of 1/2 over three decades in $\Delta\phi$. The noise in the data is likely due to a finite number of systems used for each data point. The respective values of the fitting parameters $A$ and $B$ for each separation distance $D/R$ are shown in the inset. The coefficient $A$ represents the strength of the power law dependence, and increases rapidly for small $D/R$. The exception is when the pinned particles are touching, where depletion forces are present in addition to the Casimir-like force. The constant $B$ has a direct interpretation; it is the limiting value of the force when $\Delta\phi=0$, or alternatively, when the correlation length $\xi=\infty$. $B$ also shows a sharp decrease near $D/R=0$, for the same reasons described above.

Critical Casimir forces should depend sensitively on the size of the correlation length $\xi$, which diverges near the critical point \cite{Balibar2005,Hanke1998,Gambassi2009,Krech1994,Gambassi2011}. However, as mentioned previously, the jamming transition resembles a random first-order transition \cite{birolir}, and involves the divergence of multiple length scales, some of which have only recently been illuminated in detail \cite{goodrich3,Mailman2011}. Among these a longitudinal or rigidity length scale, $l^{*}$ \cite{goodrich2, wyart}, related to the maximum size of a stable cluster within the system, and a transverse length scale, $l_{T}$ \cite{goodrich3}, that controls the stability of the system against transverse plane-wave perturbations, have been proposed. Both length scales seem to be relevant to control the stability of the jammed packings but scale differently with $\Delta\phi$. Specifically, $l^*\sim\sigma\Delta\phi^{-1/2}$ and $l_T\sim\sigma\Delta\phi^{-1/4}$. It is not clear which one of these length scales is most important for the Casimir effect in jammed systems. However, given that $l^*$ diverges more quickly, it seems logical that this length scale will play the dominant role. Thus we associate $l^*$ with the correlation length $\xi$.

With this in mind, we can write the resulting normalized force as:
\begin{align}
\left\langle \frac{ F_y}{PR}\right\rangle=B-A\frac{c\sigma}{l^*},
\label{nearjam2}
\end{align}
where $c\approx0.28$ is a numerical constant which has been recently measured in simulations of 2D frictionless disks \cite{goodrich2}. In order to tease apart the dependence of the normalized force on the remaining dimensionless parameters in Eqn.\ \ref{fulldep}, we will focus on the fitting parameters $B$ and $A$. Both will depend on $D/R$, $R/\sigma$, and $R/L$. In this case, the ``thermodynamic limit'' is achieved when $R/\sigma\rightarrow\infty$ and $R/L\rightarrow 0$. For large pinned particles, the limiting values obtained in our simulations are $R/\sigma\approx21,26,32$ and $R/L=0.083,0.105,0.13$. These parameters are coupled together because the ratio $L/\sigma$ is nearly constant for our simulations due to the fixed number of particles ($N=10,000$). Both of these parameters represent a finite-size effect that can not be ignored. However, by varying the separation $D/R$, we can accurately determine the distance-dependence of the Casimir effect. 

\begin{figure}[t]
\includegraphics[width=.47 \textwidth]{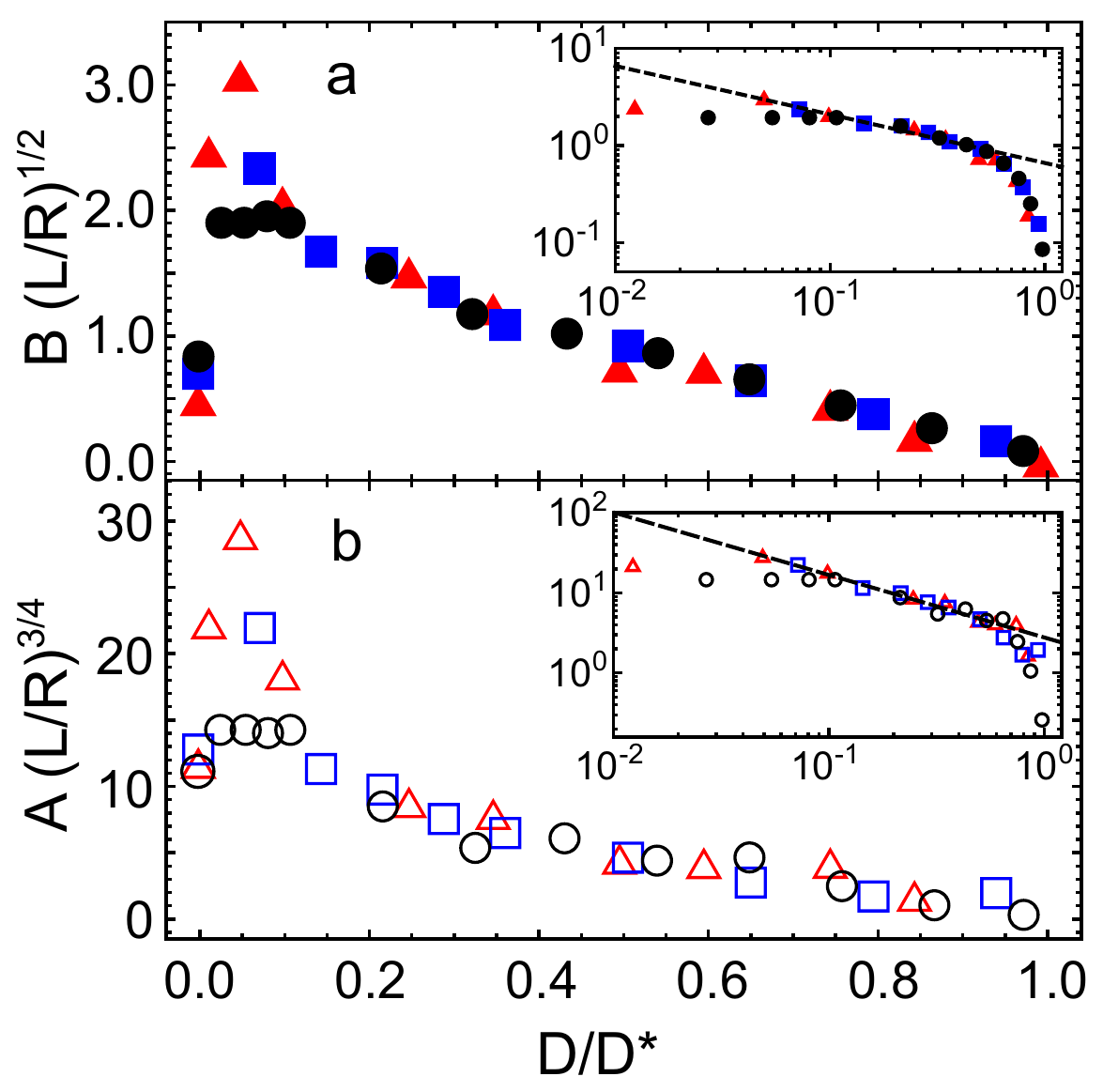}
\caption{\label{collapse} Normalized fitting parameters $B(L/R)^{1/2}$ (a) and $A(L/R)^{3/4}$ (b) versus $D/D^*$, as defined in Eqn.\ \ref{dstar}. Data is shown for three different values of $R/\sigma$: $R/\sigma=21$ ({\color{red} $\blacktriangle$}), $R/\sigma=26$ ({\color{blue} $\blacksquare$}), and $R/\sigma=32$ ({\color{black} $\newmoon$}). Open symbols are used in (b). The insets show the same data on a logarithmic scale. The dashed lines correspond to fits to the data where $D/\sigma>3.5$ and $D/D^*<6$: $B(R/L)^{1/2}\approx 0.66 (D^*/D)^{0.50\pm0.04}$ and $A(R/L)^{3/4}\approx 1.66 (D^*/D)^{0.78\pm0.05}$.
}
\end{figure}

We are able to obtain a reasonable collapse of the data by plotting $B(R/L)^{1/2}$ and $A(R/L)^{3/4}$ versus $D/D^*$, where $D^*$ is the separation at which the force must go to zero due to the symmetry of the periodic boundary conditions (Eqn.\ \ref{dstar}). The data collapse is shown in Fig.\ \ref{collapse}a-b. The collapse is poor for smaller values of $D/D^*$ because this is where $D/\sigma<5$. The scaling exponents 1/2 ($B$) and 3/4 ($A$) are a natural consequence of the dependence on $D$. If we plot the data on a logarithmic scale (insets in Fig.\ \ref{collapse}), we see that there is a small region where data is consistent with a power-law, bracketed on both sides by finite size effects which cause the data to deviate from the scaling behavior. By fitting the data in this regime, we find exponents in good agreement with the scaling obtained by collapsing the data. These fits are shown by the dashed lines in Fig.\ \ref{collapse}a-b. 

Thus in this intermediate regime, where the finite-size effects are not dominant, we find:
\begin{align}
\left\langle \frac{ F_y}{PR}\right\rangle\approx0.66\left(\frac{D^*R}{DL}\right)^{1/2}-1.66\left(\frac{D^*R}{DL}\right)^{3/4}\frac{c\sigma}{l^*},
\label{fullexp1}
\end{align}
In the limit $R/L\rightarrow 0$, the distance $D^*\rightarrow L$, and Eqn.\ \ref{fullexp1} becomes
\begin{align}
\left\langle \frac{ F_y}{PR}\right\rangle\approx0.66\left(\frac{R}{D}\right)^{1/2}-0.47\left(\frac{R}{D}\right)^{3/4}\frac{\sigma}{l^*},
\label{fullexp2}
\end{align}
where we have used the numerical value for $c$. Conveniently, as the system size $L\rightarrow\infty$, its dependence drops out of the expression for the force. Although our range of data is not very large, we may say that it is consistent with Eqn.\ \ref{fullexp2}, which gives us some expectation for the dependence of the force on the particle separation when finite-size effects are negligible. 

The scaling of critical Casimir forces is often broken into two regimes where $D\ll R$ (Derjaguin) and $D\gg R$ \cite{Hanke1998}. In both cases, the force will scale as $F\propto (R/D)^\beta \theta(D/\xi)$, where the exponent $\beta$ and the function $\theta$ are different in each regime. The universal dependence on $D/\xi$ is a defining feature of traditional Casimir forces \cite{fish}. Equation \ref{fullexp2} is not consistent with this form, so it is unclear if the Casimir-like force between the pinned particles is directly related to quantum or critical Casimir forces.  However, as we show in the next section, one of the controlling parameters for both the sign and size of all Casimir forces, namely, the boundary condition at the confining surfaces, varies significantly in jammed systems as the critical point is approached. This is in contrast to the most well-studied examples of critical Casimir forces: binary liquid mixtures near a critical point. Although the correlation length diverges as the temperature approaches the critical point, the hydrophilicity/hydrophobicity of the confining surfaces remains constant. This is perhaps the most important distinction between Casimir-like forces in jammed systems and traditional Casimir forces. 


\subsection{Origin of the Casimir Force}

The length scale $l^*$ determines the maximum size of a rigid cluster with no mechanical constraints at the boundaries \cite{goodrich2}. This argument is essentially based on balancing the excess number of contacts within a volume of jammed particles with the number of contacts lost at the free boundaries. At $\Delta\phi=0$, frictionless, jammed systems are isostatic, so there are exactly enough contacts to constrain the degrees of freedom, so $l^*\rightarrow\infty$, meaning that if one contact is broken then the whole system falls apart. Phenomenologically, this is consistent with our simulations. For large volume fractions when $l^*<D$, a rigid cluster can exist in between the pinned particles, so that it need not interact with the boundaries of the pinned particles to maintain a force balance. However, as $l^*\rightarrow \infty$ ($\Delta\phi\rightarrow 0$), the cluster can not maintain its rigidity without interacting with the boundaries of the pinned particles. 

The type of interaction will depend on the boundary conditions for the fluctuating fields in the system, as is the case in quantum and critical Casimir forces. In the case of critical Casimir forces in binary liquid mixtures near a critical point, the Casimir force can be positive or negative depending on the preference of each phase to be adjacent to the solid boundary, in effect, the hydrophobicity of the boundary \cite{bechinger}. As we will show using an alteration to the standard Maxwell counting argument for particle contacts, pinned particles in jammed systems require a reduction in the mean contact number per particle, $\langle z\rangle$. 

Let's begin by assuming an isostatic ($\Delta\phi=0$), jammed system contains $N$ free particles and $m$ identical pinned particles. Then there are $Nd$ degrees of freedom, where $d$ is the dimension of the system. This must be equal to the number of constraints, or contacts. In the bulk, there are $z_{bulk}$ contacts between free particles, and $z_{bound}$ contacts between free and pinned particles. We are assuming that the pinned particles are sparse, and do not touch. Thus $Nd=z_{bulk}+z_{bound}$. We also assume that each contact $z_{bound}$ represents a single free particle in contact with a single pinned particle, so $z_{bound}=N_{bound}$, where $N_{bound}$ is the number of free particles in contact with the boundary of a pinned particle. 

The mean number of contacts for the $N$ particles is given by $\langle z \rangle=(2z_{bulk}+z_{bound})/N$. The factor of 2 accounts for the fact that each bulk contact is shared between two free particles, but bound contacts are not. Thus we find that with the addition of $m$ pinned particles, the coordination number becomes
\begin{equation}
\langle z\rangle=2d-\dfrac{N_{bound}}{N}.
\label{zmean}
\end{equation}
Thus the mean contact number is reduced by an amount $N_{bound}/N$. Far away from a pinned particle, we might expect the bulk system to behave like any other jammed system, so the local mean contact number would be $2d$. For simplicity, let's assume that this reduction in mean contact number is a result of a surface of particles surrounding each pinned particle which has, on average, less contacts than particles in the bulk. We can estimate the thickness of this surface layer in the following way. Instead of being evenly distributed throughout the bulk, this mean contact number reduction will be confined to a layer of thickness $\Delta R$, which surrounds $m$ pinned particles. In two dimensions, the reduction in contact number in this layer, $\langle z \rangle_{layer}$, will be given by total reduction in $\langle z \rangle$ multiplied by the ratio of the total area of the system to the fraction of area included in the localized layers:
\begin{equation}
\langle z \rangle_{layer}=4-\dfrac{N_{bound}}{N}\times\dfrac{L^2-m\pi R^2}{m\pi(R+\Delta R)^2-m\pi R^2}.
\label{zlayer1}
\end{equation}

The area of the total bulk system is reduced by an amount $m\pi R^2$ due to the presence of the $m$ pinned particles. When Casimir-like forces are dominant, i.e. $R/\sigma\gg 1$, we can assume that the boundary  of the pinned particles is completely surrounded  with free particles, so that $N_{bound}\approx2\pi R m/(2\sigma)$. In addition, at jamming, the system size and particle number are related through the critical packing fraction: $L^2-m \pi R^2\approx N \pi\sigma^2/\phi_{c}$, where $\phi_{c}\approx0.84$. Thus $\langle z \rangle_{layer}$ becomes
\begin{align}
&\langle z \rangle_{layer}=4-\dfrac{\pi\dfrac{\sigma}{R}}{\phi_c\dfrac{\Delta R}{R}\left(2+\dfrac{\Delta R^2}{R^2}\right)}
\label{zlayer2}
\end{align}
In the absence of pinned particles, $\langle z \rangle_{layer}=2d=4$. However, particles need a minimum of $d-1=3$ contacts for stability. Thus we expect $3<\langle z \rangle_{layer}<4$. Near jamming for $R/\sigma=32$, we find that $\langle z \rangle_{layer}\approx3.8$. Using Eqn. \ref{zlayer2}, this suggests that $\Delta R \approx 4.4\sigma$. Thus the thickness of this region of reduced contact number is expected to be a few particles thick. We note that the thickness of the layer is independent of system size ($L$) and the number of pinned particles ($m$).

\begin{figure}[t]
\includegraphics[width=.47 \textwidth]{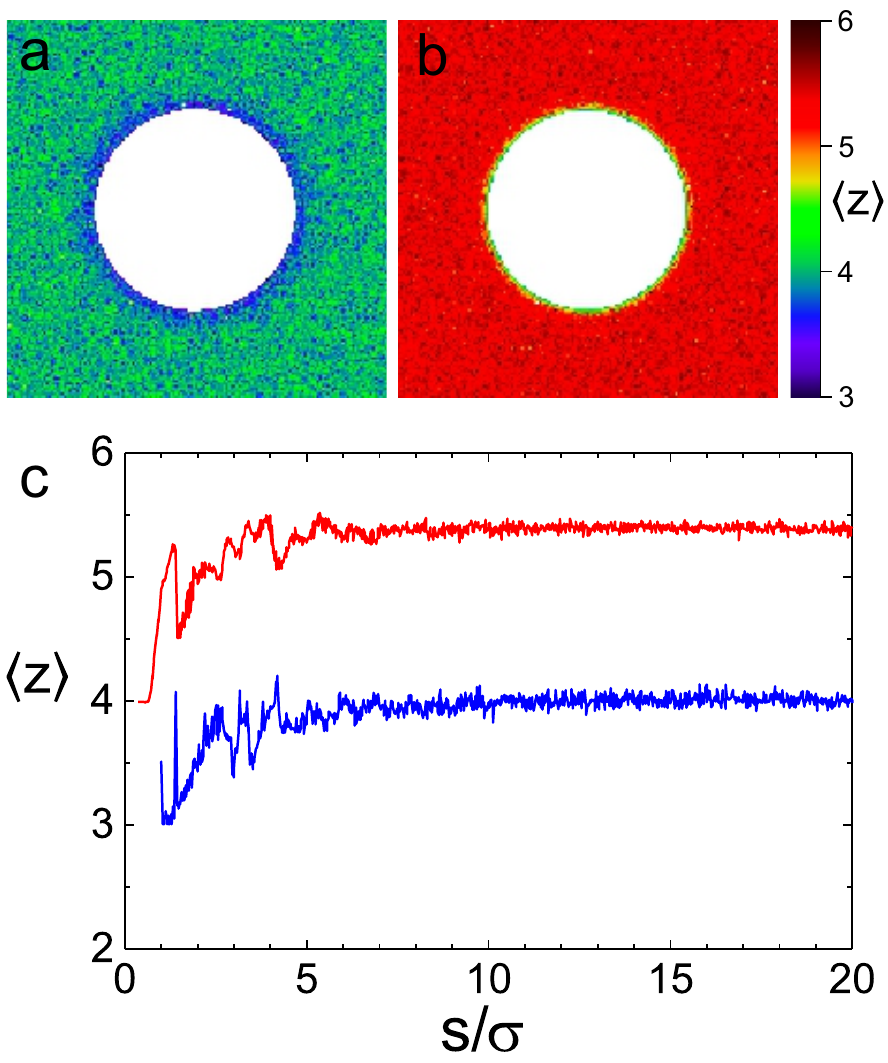}
\caption{\label{b_cond} (a-b) Coarse-grained color map of the mean contact number per particle, $\langle z \rangle$, around an isolated pinned particle. Data is shown for $R/\sigma=32$ and $D/R=1.9$, and for two different packing fractions: (a) $\Delta\phi=10^{-5}$, and (b) $\Delta\phi=0.15$. The data is averaged over 200 independent systems with $N$ = 10,000 particles each. (c) Plot of $\langle z \rangle$, averaged around the perimeter of the pinned particle, versus the normalized distance, $s/\sigma$, from boundary of the pinned particles. Data is shown for $\Delta\phi=10^{-5}$ ({\color{blue} blue line}), and $\Delta\phi=0.15$ ({\color{red} red line}). For both low and high packing fractions, $\langle z \rangle$ is reduced near the boundary of the pinned particles. Large variations near $s/\sigma=0$ are due to structural layering of particles at the boundary.}
\end{figure}

Confirmation of this prediction can been seen in Fig.\ \ref{b_cond}. Near jamming, when $\Delta\phi$ is small, the mean contact number in the bulk of the system should be $\langle z\rangle=2d$. This is the case far away from the pinned particles, where $s/\sigma\gg 1$, and $s$ is the distance between the center of a free particle and the boundary of a pinned particle. Near the boundary of the pinned particles,  $\langle z\rangle$ deviates from this value. The mean contact number decreases to $d-1=3$ at the boundary, which occurs over a few particle radii, as predicted by Eq.\ \ref{zlayer2}. This reduction in $\langle z\rangle$ near the boundary is also true at higher packing fractions, although $\langle z\rangle$ is larger since there are excess contacts in systems far from the jamming point \cite{corey}. Although not shown here, this argument can easily be extended to higher dimensions to obtain similar results by considering the volume of a shell around a $d$-dimensional sphere. 

The jamming transition is not a 2nd-order phase transition, yet it shares many similarities. There is no clear choice for a two-point static correlation function from which a growing length scale can be derived \cite{Tewari2009}. Although many quantities have been considered as order parameters for the $T=0$ jamming transition, such as geometric properties of Voronoi tessellations \cite{Morse2016} and point-to-set correlations \cite{Mailman2012}, the number of excess contacts, $\langle z\rangle-2d$, is often considered an order parameter for the system \cite{liu}. This is due to its universal scaling near the jamming transition for $d\geq2$: $\langle z\rangle-2d\propto\Delta\phi^{1/2}$. The local contact number is significantly reduced at the boundary, meaning that particles with less contacts on average will be drawn to the region between the pinned particles. However, if this region of reduced contact number is to stay in mechanical equilibrium with the rest of the system, the contacts should, on average, be stronger than contacts in the bulk of the system. This is the essential reason why the Casimir-like force is strictly positive. The boundary condition requiring a reduced number of contacts necessitates an increased strength per contact. These stronger contact forces are transmitted to the boundaries of the pinned particles, where they push them apart, resulting in a repulsive force.

\begin{figure}[t]
\includegraphics[width=.47 \textwidth]{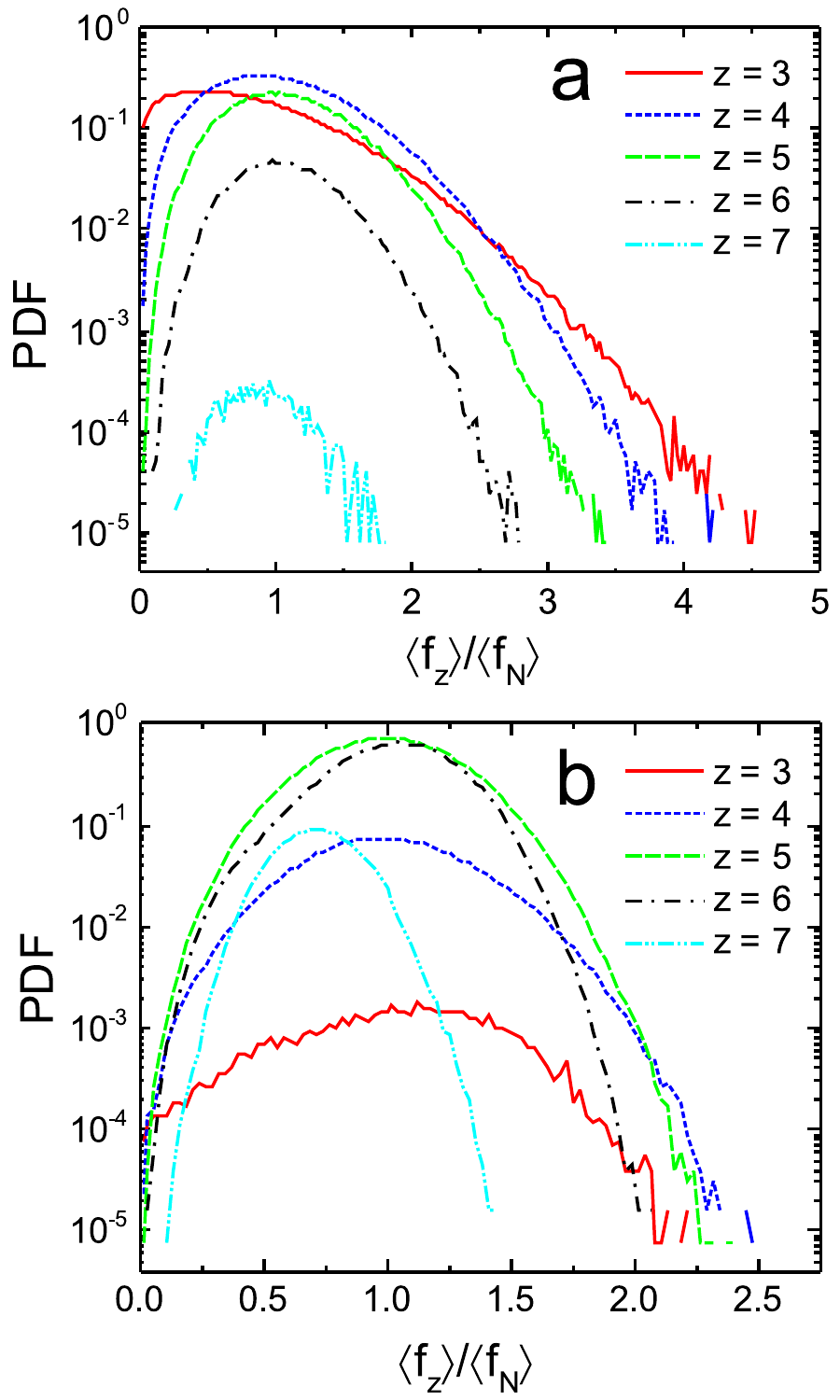}
\caption{\label{f_dist} PDFs of the mean contact force for particles with different contact numbers. The data comes from systems with no pinned particles, and is averaged over 200 systems with $N$ = 10,000 particles each. (a) $\Delta\phi=10^{-5}$. (b) $\Delta\phi=0.15$.}
\end{figure}

\begin{figure*}[t]
\includegraphics[width=.99 \textwidth]{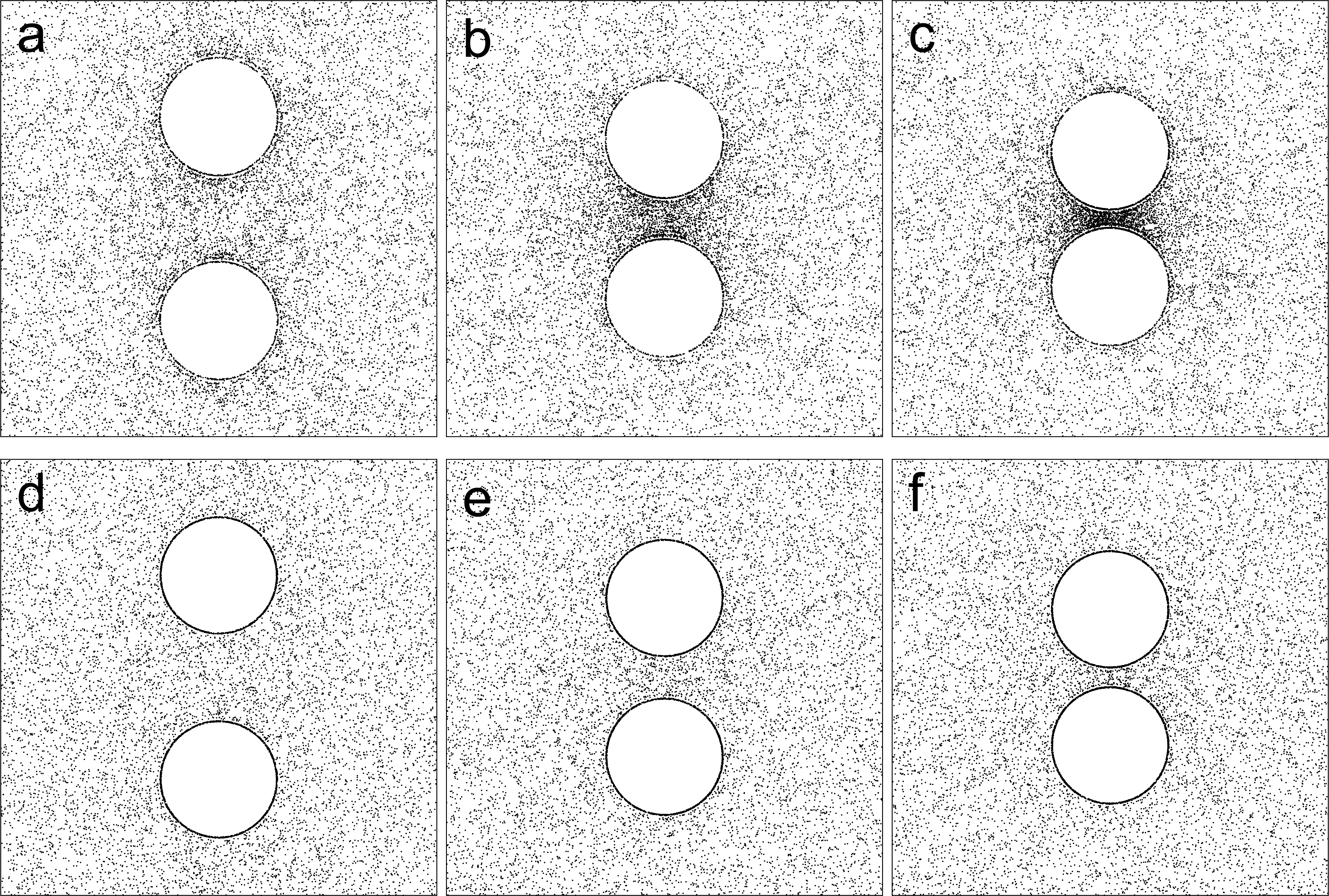}
\caption{\label{f_points} Locations of the top 0.625\% of particles with the strongest average contact forces ($\langle f_z\rangle$). For each panel, data is shown for 200 independent systems with $N$ = 10,000 and $R/\sigma=32$. (a-c) Three different values of $D/R$ (1.6, 0.8, 0.4) with $\Delta\phi=10^{-5}$. (d-f) The same particle separations with $\Delta\phi=0.15$.}
\end{figure*}

To solidify the link between reduced contact number and stronger forces, in Fig.\ \ref{f_dist} we plot the probability density functions (PDFs) of the average force per contact on a single particle with $z$ contacts:
\begin{align}
\langle f_z\rangle_i&=\dfrac{1}{z}\sum_{j=1}^{z} f_{ij},
\end{align}
where $f_{ij}$ is the magnitude of the contact force between particles $i$ and $j$, and $z$ denotes the number of contacts for particle $i$. The forces are normalized by the mean contact force on all particles in the system:
\begin{align}
\langle f_N\rangle&=\dfrac{1}{N}\sum_{i=1}^{N}\langle f_z\rangle_i.
\end{align}
We separate the PDF for all particles into separate curves for $z$ = 3, 4, 5, 6, and 7 contacts. Figure \ref{f_dist}a shows data near the jamming transition. With this normalization, the total area under all the curves in Fig. \ref{f_dist} is slightly less than unity since we have not considered $z=8$ and larger, which represents a very small fraction of the particles in the systems. An important feature to notice is that particles with the strongest contact forces only have 3 contacts. This may be expected given that these particles are in mechanical equilibrium with the rest of the system, they need stronger forces if they have fewer contacts. Figure \ref{f_dist}b shows the same PDFs at $\Delta\phi=0.15$, far from the jamming transition. For this value of $\Delta\phi$, $\langle z\rangle\approx 5.37$, the same as the data in Fig.\ \ref{b_cond}.  The distributions are more narrow with rapidly decreasing tails for strong forces, and particles with $\langle z\rangle$ = 4 and 5 contacts dominate the strongest forces. It is important to note that these distributions are taken from systems with no pinned particles, although they change very little from distributions in the presence of pinned particles. 

\begin{figure}[t]
\includegraphics[width=.47 \textwidth]{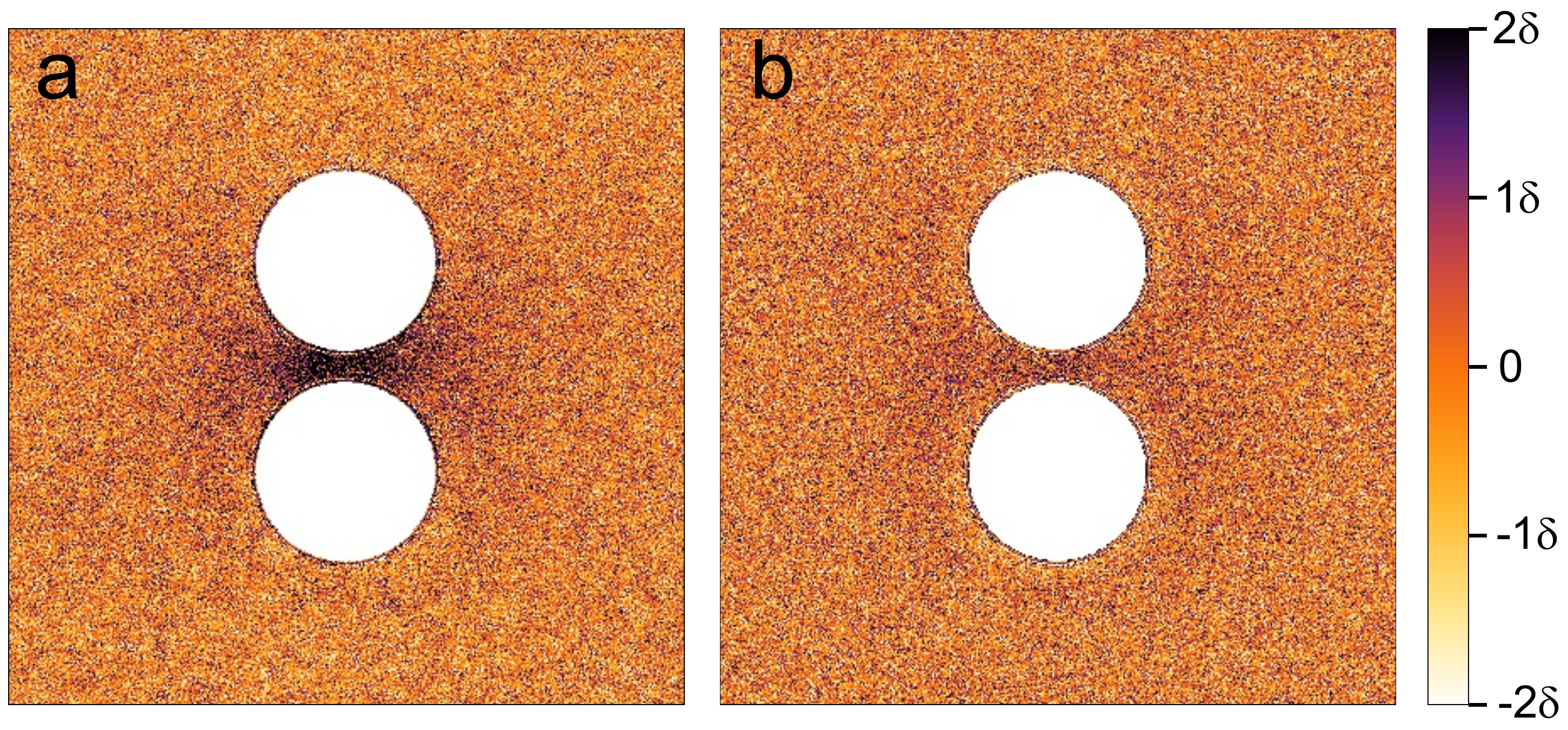}
\caption{\label{pressure} Coarse-grained map of $\langle f_z\rangle$ for $D/R=0.4$ with $\Delta \phi=10^{-5}$ (a) and $\Delta \phi=0.15$ (b). The data is averaged over 200 independent systems with $N$ = 10,000 and $R/\sigma=32$. The colors represent mean force ({\color{orange} orange}) and number of standard deviations, $\delta$, from the mean force. }
\end{figure}

As stated above, the strongest forces should be localized to the region between the pinned particles. In order to visualize this localization, we look for where the particles with the strongest average contact force are located. Figure \ref{f_points} shows the strongest 0.625\% of particles for 200 independent systems for two different values of $\Delta\phi$, corresponding to Fig.\ \ref{f_dist}, as well as different values of $D/R$. At $\Delta\phi=10^{-5}$ (panels a-c), the data represents particles with $\langle f_z\rangle/\langle f_N \rangle\gtrsim 2.8$, which mostly consists of particles with $z$ = 3 contacts. We find that there is a clear tendency for strong forces to localize in between the pinned particles, especially when the pinned particles are close. This average localization gives rise to the repulsive Casimir effect, which is prominent near jamming. When $\Delta\phi=0.15$, there are strong forces localized to the immediate boundary of the pinned particles, but this localization does not extend far into the region between the particles. The points shown in panels d-f correspond to $\langle f_z\rangle/\langle f_N \rangle\gtrsim 1.6$, where the forces are dominated by particles with $z$ = 4, 5, and 6 contacts. The net effect of this clustering can be seen in Fig.\ \ref{pressure}, which shows a color map of $\langle f_z\rangle$ for closely-spaced particles. Near the jamming transition (Fig.\ \ref{pressure}a), the average contact force in between the particles can be more than 2 standard deviations away from the mean force in the system. This deviation is significantly reduced for larger packing fractions, i.e. far from the jamming transition.

Taken together, we interpret the data from Figs.\ \ref{f_dist}, \ref{f_points}, and \ref{pressure} in the following way. Fluctuations in the particle contact number and contact strength occur throughout the bulk, and are larger near jamming, given that the length scale $l^*$ diverges \cite{wyart,goodrich2} and the PDFs in Fig.\ \ref{f_dist} have long tails. The spectrum of fluctuations must be consistent with the boundary condition at the pinned particles. The fixed boundaries of the pinned particles produce a reduction in the mean number of contacts per particle, $\langle z\rangle$. We find that this reduction is localized near the boundary of the pinned particles. As two pinned particles approach one another, the localized layers of reduced $\langle z \rangle$ begin to overlap, resulting in a repulsive interaction due to stronger contacts for particles with $z$ = 3 and $z$ = 4. However, the amplitude of this reduction has an upper bound: $\langle z\rangle$ can not drop below 3 contacts per particle in order to maintain rigidity. Thus, near jamming, the reduction in contact number is spread over a longer distance, i.e. the layer near the boundary is thicker. At larger packing fractions, $\langle z\rangle$ is larger on average, which allows for a sharper decrease in $\langle z\rangle$ near the boundary, and a thinner interaction range for the Casimir effect. Essentially, the excess contacts screen the interaction between the pinned particles. 

One important question suggested by our results, and in particular by Fig.\ \ref{b_cond}, is what happens when the pinned particles are allowed to relax with the rest of the system? How would this reduction in contact number near the surface change? We have tested this by allowing the pinned particles to relax to a zero net force, starting from a condition where they were initially pinned. On average, the pinned particles always move further apart due to the repulsive Casimir-like force, before reaching an equilibrium state (zero net force). However, when $R/\sigma\gg1$, the contact number distribution near the boundary is still identical to that of Fig.\ \ref{b_cond}. This is essentially because Eq.\ \ref{zmean} does not require that the particles be pinned. We can see this by considering that large particles immersed in a sea of smaller particles give rise to many contacts ($N_{bound}$), yet they only bring $d$ degrees of freedom per particle to the counting statistics. In this case, where the $m$ pinned particles are allowed to move, we arrive a different expression for the average contact number in the bulk:
\begin{equation}
\langle z\rangle=2d-\dfrac{N_{bound}}{N}+\dfrac{2 d m}{N}.
\label{zmean2}
\end{equation}
If $R/\sigma\gg 1$, so that each large particle has many smaller particles in contact with it boundary ($N_{bound}\gg m$), Eq.\ \ref{zmean2} reduces to Eq.\ \ref{zmean}. If the $m$ particles are the same size as the rest of the jammed particles, then they have the same number of average contacts as any other particle ($N_{bound}=\langle z \rangle m$), and Eq. \ref{zmean2} reduces to $\langle z\rangle=2d$. Thus, there will be a reduction in the number of contacts, which is still localized to the boundary of the large particles, even if the pinned particles are allow to move. 

\begin{figure}[t]
\includegraphics[width=.4 \textwidth]{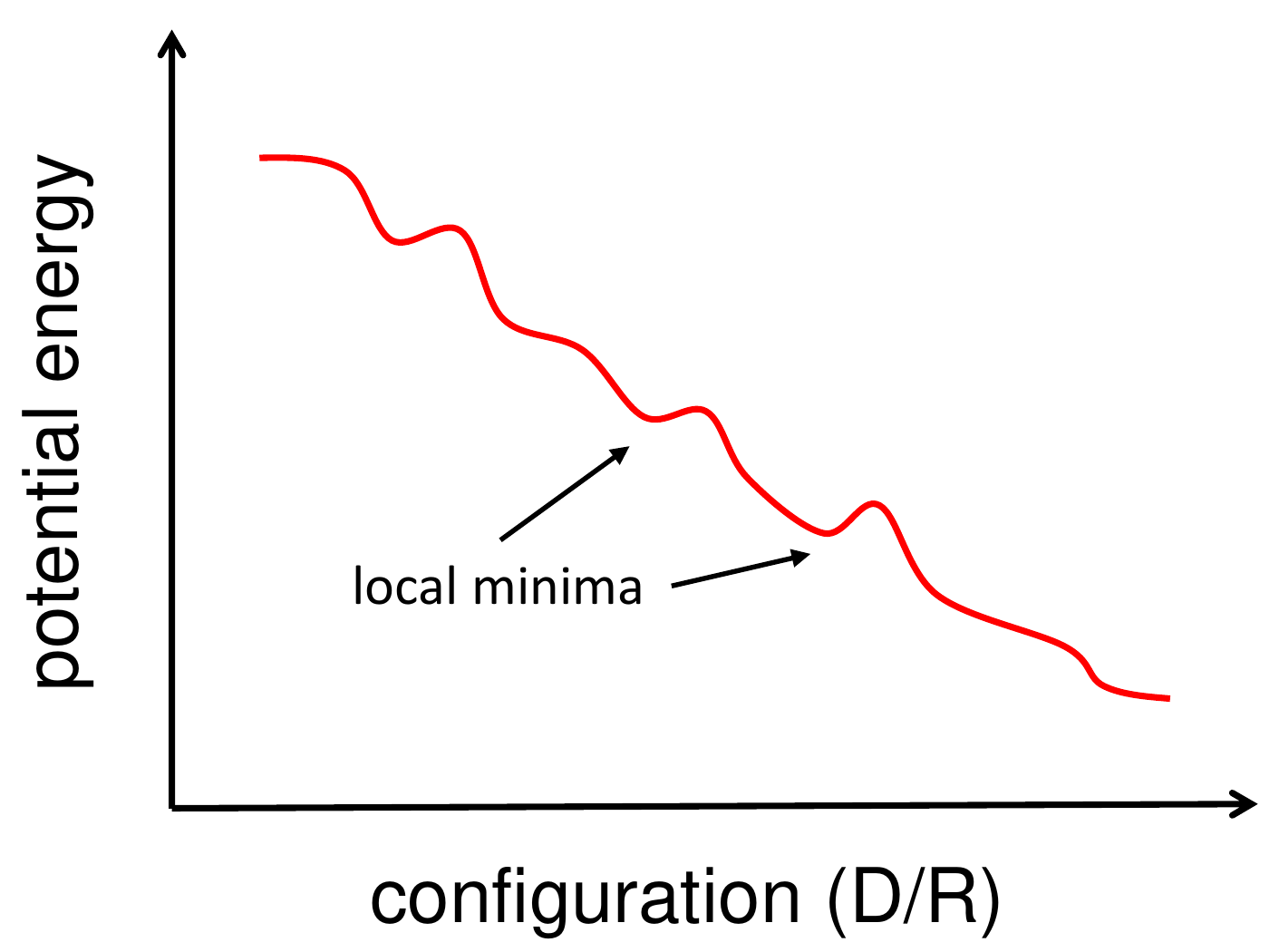}
\caption{\label{landscape} Schematic of the potential energy landscape associated with the position of the large, pinned particles.}
\end{figure}

This idea can be seen in Fig.\ \ref{landscape}, which shows an idealization of the potential energy landscape which may be associated with the separation of the large, pinned particles. In general, increasing the separation will decrease the potential energy, so that the ensemble average force between the pinned particles is always repulsive. However, for a single system, if the pinned particles are allowed to relax, they can be quenched to a small, local potential energy minimum. Given some type of excitation, perhaps thermal or an external vibration of the system, the larger particles can escape this local minimum and continue to move further apart. In jammed systems, our results imply that large particles will always tend to move away from each other, or from the fixed boundaries of the system. This may have important implications for the well-known Brazil nut effect, where the largest particles in a system end up near the free surface (away from the bottom boundary of the container) in a driven, granular system. 

It is unclear at this point how these results may be related to numerous studies investigating the force on an intruder in a granular medium \cite{Stone2004,Stone2004nat,Candelier2010,Reichhardt2010,Ding2011}. In particular, Stone et al. measure a sharp increase in the total force on an intruder approaching a solid wall. They find that the force is repulsive and depends exponentially on the distance from the wall. A direct comparison of our data with the experiment is complicated by particle-particle, particle-wall, and particle-intruder friction, as well as the free surface of the granular bed, all of which have a potent effect on the experimental measurements. In addition, in our simulations, the force for each value of $D$ is measured by fixing $D$, then quenching the smaller mobile particles. Since we have shown that the preparation of the jammed packing can affect the force, we may expect different results if we first quenched the mobile particles at a large value of $D$, then slowly brought the pinned particles together. We hope to address these questions in future work.

\section{Conclusions}

The jamming transition has been investigated for nearly two decades. We now know that it is a unique example of a random first order transition, and shares similarities with the glass transition. It shares properties with both first and second order phase transitions, such as a finite jump in the order parameter, power law scaling of system properties, diverging correlation lengths, and strong fluctuations near the critical point ($\Delta\phi=0$). The latter two are the main ingredients necessary for macroscopic Casimir forces. Here we have shown how Casimir-like forces arise between two pinned particles in an ambient jammed system of frictionless particles. This force is an ensemble average over many realizations of jammed packings. When the pinned particles are small, they experience sharp fluctuations in their interaction force due to the local structure of the packing, as defined by the pair distribution function. However, at large separations, the particles still retain a small repulsive force.

When the pinned particles are much larger than the ambient jammed particles, the long-ranged Casimir-like force is dominant, and is purely repulsive. Near jamming, the Casimir-like force obeys a universal scaling law, $A-B\Delta\phi^{1/2}$, however the exact values of $A$ and $B$ depend on the preparation of the system. Although we do not currently have an explanation for the exponent value, 1/2, we suspect that this is related to the well-known 1/2 exponent seen in the behavior of the number of excess contacts versus $\Delta\phi$ since the origin of the Casimir effect stems from fluctuations in contact number. The force also depends on the distance between the pinned particles, and is proportional to $(R/D)^{1/2}$ at the jamming transition (Eq.\ \ref{fullexp2}). The repulsive nature of the force and its increased magnitude at short distances can be explained by the confinement of contact number fluctuations. The boundary of the pinned particles requires a decrease in the mean contact number. Regions of particles with fewer contacts have stronger contacts on average, giving rise to an increase in pressure between the pinned particles.

Casimir forces are intimately connected to system fluctuations. Both quantum and thermal fluctuations are easier to define, whereas ensembles of jammed systems depend on the user protocol for generating them. This makes any attempt at an analytic theory more complicated, however, we speculate that the Casimir effect in jammed systems may be intimately related to those found in other nonequilibrium systems \cite{Hanke2013,Brito2007,grano2,Larraza1998,Obukhov2005,catt,duncan,zuri,villa,reza,Ray2014,Kirkpatrick2015,Najafi2004}. Thus we do not expect Casimir-like forces near the jamming transition to follow the same scaling as quantum or critical Casimir forces. Unlike classical critical behavior, the jamming transition is a rare example of a random first-order transition which involves two equally-important diverging length scales, and the boundary condition at interfaces (contact number and contact strength) vary as $\Delta\phi\rightarrow 0$. Nevertheless, there may be some similarities with traditional Casimir forces. The correlation length $l^*$ is associated with the onset of floppy modes which have been shifted upwards in frequency when $\Delta\phi>0$ \cite{liu}. Near jamming, the excess low-frequency modes may serve the same role as Goldstone bosons generated by symmetry breaking in critical Casimir forces. In order to verify this potential hypothesis and fully map out the scaling dependencies of the Casimir-like force on the particle separation, radius, and system size, a more extensive set of simulations would be necessary with many more jammed particles ($N\gg10,000$). 

Although our simulations are strictly two-dimensional, Eq.\ \ref{zlayer2} can be derived in any dimension, given knowledge of the critical packing fraction $\phi_c$. Thus we expect our results to be qualitatively valid in three dimensions, although the exact dependence on parameters such as $D/R$ may change.  It also remains to be seen how these results extend to finite temperature, especially since a full understanding of the jamming transition at finite temperature is complex, and a subject of active research \cite{DeGiuli2015,Caswell2013,Ikeda2013}. Figure \ref{back_and_forth} would suggest that even upon some form of annealing, the Casimir effect still remains. Also, Fig.\ \ref{landscape} suggests that if the large particles are allowed to move, then they would continue to move farther apart until the system can not escape from a local minimum. We leave these questions open for further study.

Finally, the addition of friction may play a crucial role in the behavior of the Casimir effect, since it strongly affects the average contact number per particle \cite{vanhecke}. Generically, there are still excess contacts above the jamming transition in frictional systems, which would act to ``screen'' the Casimir-like force, as shown in Fig.\ \ref{f_points}. We suspect that a repulsive, Casimir-like force would still dominate near jamming, although it is unclear if the universal behaviors would remain since the jamming transition is more well-defined in the absence of friction \cite{Bi2011}. Although simulations allow precise control over the particle interaction and ensemble generation, it remains to be seen if these ideas can be experimentally observed. Our lab is currently developing a series of ongoing experiments to test these theoretical predictions. 

\section{Acknowledgments}

We would like to that Carl Goodrich, Wouter Ellenbroek, Eric Corwin, and Sidney Nagel for helpful discussions. This work was supported by the National Science Foundation through grant NSF DMR-1455086.

\end{document}